\documentclass[prd,aps,twocolumn,superscriptaddress,nofootinbib,floatfix]{revtex4-2}
\usepackage{graphicx}
\usepackage{epsfig}
\usepackage{bm}
\usepackage{amssymb}
\usepackage{amsmath}
\usepackage{mathrsfs}
\usepackage{color}
\usepackage{float}
\usepackage{setspace}
\usepackage{ulem}

\begin{document} 

\author{Shou-shan Bao}
\email{ssbao@sdu.edu.cn}
 \affiliation{\normalsize\it Key Laboratory of
Particle Physics and Particle Irradiation (MOE),Institute of
Frontier and Interdisciplinary Science, Shandong University,
(QingDao), Shandong 266237, China }

\author{Wenhai Gao}
\email{gaowenhai312@gmail.com}
 \affiliation{\normalsize\it Key Laboratory of
Particle Physics and Particle Irradiation (MOE),Institute of
Frontier and Interdisciplinary Science, Shandong University,
(QingDao), Shandong 266237, China }

\author{Hong Zhang}
\email{Hong.zhang@sdu.edu.cn}
 \affiliation{\normalsize\it Key Laboratory of
Particle Physics and Particle Irradiation (MOE),Institute of
Frontier and Interdisciplinary Science, Shandong University,
(QingDao), Shandong 266237, China }

\author{Jian~Zhou}
\email{jzhou@sdu.edu.cn}
 \affiliation{\normalsize\it Key Laboratory of
Particle Physics and Particle Irradiation (MOE),Institute of
Frontier and Interdisciplinary Science, Shandong University,
(QingDao), Shandong 266237, China }
\affiliation{\normalsize\it Southern Center for Nuclear-Science Theory(SCNT), Institute of Modern Physics, Chinese Academy of Sciences, HuiZhou, Guangdong, 516000, China}

\title{Constraining axion-gluon coupling in monohadron processes}

\begin{abstract}
The axion-gluon coupling can be constrained directly through hard exclusive processes at the LHC. Specifically, we study the associated production of a long-lived axion with a $\rho^0$ meson in ultraperipheral $AA$ collisions and in $pp$ collisions. With the axion escaped from the detector, the final state is characterized by a monohadron signature. The main background in our analysis originates from the $\rho^0+\pi^0$ process, where the photons from the $\pi^0$ decay are undetected due to limited detector performance. Our analysis yields an exclusion limit of the axion-gluon coupling  that is comparable to the limit obtained from the monojet process at the LHC.
\end{abstract}
\maketitle

\section{Introduction}
While the Standard Model (SM) has achieved remarkable success, the existence of dark matter suggests that new physics lies beyond its boundaries. Among the candidates for light dark matter, the axion stands out  for various reasons. Originally proposed to solve the strong $CP$ problem within the Standard Model \cite{Peccei:1977hh, Peccei:1977ur, Weinberg:1977ma, Wilczek:1977pj, Kim:1979if, Shifman:1979if, Zhitnitsky:1980tq, Dine:1981rt}, axions and axionlike particles (ALPs) may emerge in various extensions of the SM. ALPs are $CP$-odd, pseudo-Nambu-Goldstone scalar particles resulting from spontaneously broken symmetries at a high-energy scale denoted as $f_a$. These particles remain singlets under the SM gauge interactions. Their intriguing properties connect them to multiple open questions in both particle physics and astrophysics \cite{Chang:2000ii, Boehm:2014hva, Berlin:2014tja, Krasznahorkay:2015iga, Marciano:2016yhf, Feng:2016ysn, Ellwanger:2016wfe, Han:2022kvj,Hui:2016ltb,Hu:2000ke,Chakraborty:2024tyx}.  

Depending on various motivations, the proposed mass of the ALPs could span an impressive range, from $10^{-21}$eV up to the TeV scale.  For TeV scale ALPs, their potential to decay into dijets or dielectroweak gauge bosons allows for direct detection at high-energy colliders~\cite{Han:2022mzp,Bao:2022onq}. However, in the mass range between 1 MeV and 1 GeV, ALP-photon and ALP-gluon interactions become pivotal for designing a detection strategy. Numerous experiments have probed such parameter space~\cite{Dobrich:2015jyk,Balkin:2023gya,Ferber:2022ewf}. For light ALPs with masses below MeV, their sole decay channel is into photons, with the decay width scaling as $m_a^3$. Consequently, these light ALPs are typically long-lived, traversing considerable distances before decaying. Detection strategies include mono-X(X=photon, $Z/W$, jet, Higgs) processes and nonresonance production at both leptonic and hadronic colliders \cite{Dolan:2017osp,Brivio:2017ije,ATLAS:2024qoo}. Additionally, long-lived ALPs can be constrained by studying the invisible decay of the $Z$ boson, the Higgs boson, and hadrons~\cite{Bauer:2017ris,CLEO:1994hzy,BaBar:2010eww,Bertholet:2021hjl,DiLuzio:2024jip,Li:2024thq,Liu:2023bby,Zhang:2023wlt,Wang:2024tre,Chakraborty:2021wda}. Beyond collider experiments, cosmological and astrophysical observations impose further constraints. Big bang nucleosynthesis, cosmic microwave background distortions, and extragalactic background light measurements exclude significant portions of the parameter space for the case of small ALP-photon couplings~\cite{Millea:2015qra,Chang:2018rso,Davoudiasl:2019nlo,Cadamuro:2011fd,Ertas:2020xcc}. Furthermore, ultralight ALPs can form clouds around Kerr black holes via superradiance~\cite{Brito:2015oca}, providing a unique avenue for testing ALP properties through black hole observations~\cite{Brito:2014wla,Chen:2021lvo,Ng:2020ruv,Cheng:2022jsw,Guo:2022mpr,Brito:2017zvb}. Although the motivation for the QCD axion and the ALPs is very different, the strategies for studying their properties are very similar. In this paper, we refer to both QCD axion and ALPs as axion for brevity.

In this work, we investigate the possibility of constraining the axion properties via diffractive scattering, in which the initial particles remain intact after the interaction. These processes have been widely studied to extract the general parton distributions (GPDs) (for reviews and  discussions about the factorization property, see \cite{Goeke:2001tz,Diehl:2003ny,Belitsky:2005qn,Boffi:2007yc,Qiu:2022bpq,Qiu:2022pla}). Compared to the inclusive production of axions, the SM background can be further suppressed using the parity of the final states in the diffractive processes. Take the joint production of an axion with a $\rho^0$ in photon-nucleus diffractive collision as an example. With the axion evading detection, the final state presents a special monohadron signature. 
Because of the limited detector efficiency, the SM background is the joint production of a $\rho^0$ with a $\pi^0$, followed by complete undetection of both photons from the $\pi^0$ decay.

The incoming photon interacts with the nucleus by splitting into a quark-antiquark pair, commonly known as a color dipole.
The ratio of dipole-nucleus elastic cross section to the total cross section increases with the center-of-mass energy $\sqrt{s}$, approaching 1/2 in the $\sqrt{s}\to \infty$ limit, which is usually referred to as the ``black-disk limit" \cite{Frankfurt:2005mc}. To  maximally enhance the high-energy photon and gluon fluxes, we study the ultraperipheral collisions (UPCs) between two heavy nuclei. The UPC processes has been widely used to study nuclear structure~\cite{Xiao:2020ddm,Xing:2020hwh,Zha:2020cst,Brandenburg:2021lnj,Hagiwara:2020juc,Hagiwara:2021xkf,Lin:2022flv,Wang:2022gkd,Zhou:2022gbh,STAR:2022wfe,Zhao:2023nbl,Li:2023yjt,Xie:2023vfi,Shi:2023nko}, exotic hadron states~\cite{Niu:2022cug}, and new physics via novel polarization-dependent observables in UPCs~\cite{Xu:2022qme,Shao:2023bga}. In 
 UPCs, the photon flux is enhanced by $Z^2$ and the gluon flux is amplified by $A$, where $Z$ and $A$ are the atomic number and the mass number, respectively. Comparably, the proton-proton hard exclusive process lacks such enhancements, but its accumulated luminosity is much higher. In this work, we investigate the potential of both processes in determining the exclusion limit of the axion coupling.

Our goal is to constrain the axion-gluon coupling. It can be indirectly explored through the axion-hadron couplings with the help of chiral perturbation theory \cite{GrillidiCortona:2015jxo,Bauer:2017ris,Aloni:2018vki}, or directly in the monojet searching at the LHC \cite{ATLAS:2021kxv}. In this work, we demonstrate that this coupling can also be straightforwardly measured in the associated production of an axion with a single $\rho^0$, with the transverse momentum of the $\rho^0$ being more than a few GeV. The axion can be radiated from the initial photon, the quarks, or the exchanged gluons (see Fig.~\ref{rhoa}). The diagrams with the axion connecting to the quarks are suppressed by $m_q/l_{\rho\bot}$, with $m_q$ the light quark mass and $l_{\rho\bot}$ the transverse momentum of the $\rho^0$. Consequently these diagrams can be safely disregarded in high-energy events. As a result, we effectively work with the Lagrangian of the axion,
\begin{align}\label{eq:L}
\mathcal{L} = \frac{1}{2}\partial_\mu a \partial^\mu a - C_{a\gamma\gamma}\frac{a}{f_a}F_{\mu\nu}\widetilde{F}^{\mu\nu} 
- \frac{a}{f_a}G_{\mu\nu,c}\widetilde{G}^{\mu\nu,c},
\end{align}
where $\widetilde{F}^{\mu\nu}=\epsilon^{\mu\nu\lambda\tau}F_{\lambda\tau}/2$ and similarly for $\widetilde{G}^{\mu\nu,c}$. 
In this Lagrangian, the axion-gluon coupling is normalized and $C_{a\gamma\gamma}$ is then the ratio of the axion-photon to the axion-gluon coupling.
Note the coefficient of the last term in Eq.~\eqref{eq:L} is different in literatures, depending on the definition of $f_a$.
This coefficient is usually chosen as $\alpha_s/(8\pi f_a)$ for QCD axion and $c_g/f_a$ for ALPs. 
The constraints from different $f_a$ definitions can be simply related by multiplying overall factors. Axions with mass $m_a\lesssim 500$~MeV decay dominantly to two photons, with the effective coupling \cite{Bauer:2017ris,Aloni:2018vki}
\begin{align}
C_{a\gamma\gamma}^\text{eff} \approx C_{a\gamma\gamma} &-(1.92\pm 0.04)\frac{\alpha_\text{em}}{\alpha_\text{s}}\nonumber\\
&-\frac{m_a^2}{m_\pi^2-m_a^2}\frac{m_d-m_u}{m_d+m_u} \frac{\alpha_\text{em}}{\alpha_\text{s}}.\label{eq:effphoton}
\end{align}
The monohadron analysis is applicable only in the case where the axion is long-lived, which implies 
that this coupling must be small. Then, as long as $m_a$ is not close to $m_\pi$, it requires
\begin{align}
C_{a\gamma\gamma}\approx 1.92 \frac{\alpha_\text{em}}{\alpha_\text{s}},\label{eq:cphoton}
\end{align}
which is used in the calculation below.

In addition to $\rho^0$, the $\omega$ and $\phi$ mesons can also be produced in the same process. 
Since the production cross section of $\rho$ is larger than others, we only study the joint production of $\rho^0$ and a long-lived axion to investigate the power of the diffractive process in constraining $f_a$ for brevity. Including more hadrons can improve the constraint on $f_a$.  At low energy the axion-meson interaction can also originate from the Wess-Zumino-Witten term~\cite{Wess:1971yu,Witten:1983tw}, which plays an important role in low-energy physics.
Such interactions are explored in Ref.~\cite{Harvey:2007rd,Harvey:2007ca,Bauer:2020jbp,Bai:2024lpq}. Recently, the photoproduction of axion induced by such terms is studied in Ref.~\cite{Chakraborty:2024tyx}. In this work, we do not include the Wess-Zumino-Witten term.

We perform the calculation within the well-established QCD collinear factorization framework, in which the scattering amplitude is expressed as the convolution of the hard coefficient function, coherent photon distribution, gluon GPD, and the distribution amplitude (DA). The process can be visualized as follows: the incoming quasireal photon emitted from one of the protons or nuclei fluctuates into a quark-antiquark pair, followed by a coherent scattering with another proton or nuclei through two gluon exchanges. The axion is emitted from the initial photon or one of the exchanged gluons. The produced $\rho^0$ meson then acquires large transverse momentum via the recoil effect. The initial protons or nuclei remain intact after collisions ensuring low background rates. Our numerical estimations suggest that such a collider search could provide an exclusion limit comparable to the monojet search at the LHC.

The paper is structured as follows. We first carry out the calculation of hard-scattering amplitude and present the analytical result of the $\rho+a$ production in Sec.~\ref{sec:signal}. The SM process $\rho+\pi$ is discussed in Sec.~\ref{sec:background}. In Sec.~\ref{sec:constraint}, we present the numerical results and study how this process sets bounds on the axion-gluon coupling. 
Finally, we summarize our result in Sec.~\ref{sec:summary}.

\section{Exclusive photoproduction of $\rho^0+a$}\label{sec:signal}

In this section, we first calculate the exclusive photoproduction rate of $\rho^0+a$ in the ultraperipheral heavy ion collisions. We consider a linearly polarized quasireal photon from one nucleus colliding with two gluons from the other nucleus,
\begin{eqnarray}
\gamma + {\cal A} \rightarrow \rho^0 + a + {\cal A},
\end{eqnarray}
where ${\cal A}$ denotes the nucleus.
The coherent photon luminosity from a nucleus gives a $Z^2$ enhancement, where $Z$ is the nuclear atomic number.
The gluon luminosity is also amplified by a factor of the mass number $A$.
Therefore the production rate of axion with a $\rho$ meson is enhanced by a factor of $Z^2 A$ compared to $p p$ collisions.
One could obtain an estimate of the production rate in $pp$ collisions by setting $Z=A=1$.
Nonetheless, we employ the more accurate photon parton distribution fucntion (PDF) if the incoming photon is from a proton. 

\begin{figure}[!htpb]
\centering
\includegraphics[width=0.45\textwidth]{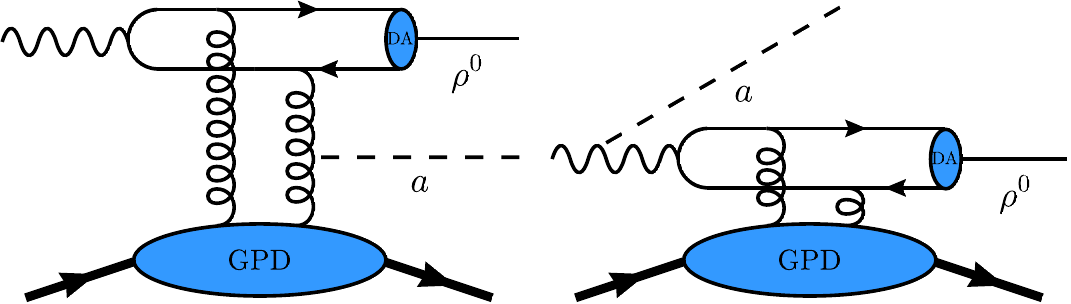}
\caption{Sample Feynman diagrams contributing to exclusive $\rho^0+a$ production.}\label{rhoa}\end{figure}

The axion could couple to both gluons and quarks in the nucleus. 
The amplitudes with the axion attaching to the quarks are proportional to the quark mass, which could be safely ignored in high-energy processes, making this process ideal to directly probe the coupling between the axion and gluon.
There are in total 18 Feynman diagrams similar to the ones showed in Fig.~\ref{rhoa} which are generated and evaluated with FeynArts and FeynCalc packages in our work \cite{Hahn:2000kx, Mertig:1990an, Shtabovenko:2016sxi, Shtabovenko:2020gxv, Shtabovenko:2023idz}.
 The nonperturbative nucleus and $\rho^0$ states are described by the GPD and the DA, respectively. Moreover, we ignore the total transverse momentum of the two gluons, which is at the order of $\Lambda_\text{QCD}$.
Then the produced $\rho^0$ and axion are back-to-back in the transverse directions. 
This assumption is valid as long as their transverse momenta are larger than $\Lambda_\text{QCD}$.

The proton GPD is defined as~\cite{Ji:1998pc}
\begin{widetext}
\begin{align}
&\int \frac{d\zeta^-}{2\pi} e^{ix_g P^+\zeta^-}\langle P'| F^{+\mu}(-\frac{\xi}{2}) F_{+\mu} (\frac{\xi}{2})|P \rangle   =\frac{1}{2} \bar u(P')\!\! \left [ \gamma^+ H^{(p)}_g(x_g,\xi,t) + \frac{i\sigma^{+\alpha} \Delta_\alpha}{2 M_p}  E^{(p)}_g(x_g,\xi,t) \right ]\! u(P),
\end{align}
\end{widetext}
where $M_p$ is the proton mass, and $P$ and $P'$ are the initial- and final-state proton momentum, respectively.
The momentum transfer is $\Delta=P'-P$ and its invariant square is $t=\Delta^2$.
The two gluons possess longitudinal momentum $(\xi-x_g)P^+$ and $(\xi+x_g)P^+$, respectively.
In the small $\xi$ limit, the contribution of $E_g^{(p)}$ can be safely ignored and one only needs to consider the function $H^{(p)}_g$, as they enter the cross section as an combination $H^{(p)}_g-\frac{\xi^2}{1-\xi^2}E_g^{(p)} $.
In the leading-log approximation, it can be parametrized in terms of the ordinary gluon PDF $G(x)$ as~\cite{Goloskokov:2005sd,Goloskokov:2006hr}
\begin{eqnarray}
H_g^{(p)}(\xi,\xi ,t)= 2\xi e^{t B/2} G(2\xi),
\end{eqnarray}
with the parameter $B=6~\text{GeV}^{-2}$  from fitting the HERA data~\cite{ZEUS:1999ptu}. 
Neglecting the shadowing effect, the nucleus GPD equals the proton GPD multiplied by the nuclear mass number $A$, i.e., $H_g^{(A)}=A\times H_g^{(p)}$.

In this work, we describe the transition from quark-antiquark pair to a longitudinally polarized $\rho^0$ with the DA of the $\rho^0$, $\Phi_{\rho} (z)$, given as~\cite{Ball:1996tb}
\begin{eqnarray}\label{eq:rho_DA}
\phi_{\rho}(z)=\frac{1}{ f_\rho}\int \! dx^- e^{iz  P\cdot x} \langle 0 | \bar \psi(0) n\!\!\!/ \psi(x)|\rho^0(P) \rangle, 
\end{eqnarray}
where $z$  denotes the momentum fraction of a light meson carried by the quark and $f_\rho=0.216$~GeV \cite{ParticleDataGroup:2022pth} is the decay constant of the $\rho^0$ meson. We further adopt the asymptotic distribution amplitude
$\phi_{\rho}(z)=6z(1-z)$
in our calculation as done in Ref.~\cite{Goloskokov:2005sd}.

With these nonperturbative matrix elements,  the sum of all Feynman diagrams gives 
\begin{align}
 {\cal M}= \frac{\vert\vec \epsilon_{\gamma \perp} \! \times \!  \vec l_{\rho \perp}\vert}{l_{\rho \perp}^2 } \!\! \int \! dx_g dz {\cal T} (l_{\rho \perp},y,\xi,x_g,z) \phi_{\rho}(z) \nonumber\\
  \times \frac{ H^{(A)}_g(x_g,\xi,t)}{(x_g-\xi +i\epsilon)(x_g+\xi-i\epsilon )},\label{eq:amp}
\end{align}
where $\mathcal{T}$ is the hard-scattering amplitude, $l_\rho$ is the momentum of the outgoing $\rho^0$ and $y$ is the longitudinal momentum fraction of the incident photon carried by the axion.
The correlation between photon polarization vector $\vec \epsilon_{\gamma \perp}$ and $\vec l_{\rho \perp}$ is expected from the parity-violating nature of the axion-gluon coupling. 

This scattering amplitude satisfies the local dispersion relation~\cite{Goloskokov:2005sd,Goloskokov:2006hr},
\begin{align}
 {\cal M}=\left[i-\frac{\pi}{2} \frac{\partial }{\partial \ln x}x\right] {\text{ Im}} {\cal M}.\label{eq:dispersion}
\end{align}
The calculation of the imaginary part is much simplified by making use of the identity
\begin{align}
\frac{ 1}{x_g-\xi+i\epsilon}={\text P} \frac{ 1}{x_g-\xi} -i\pi \delta(x_g-\xi).
\end{align}
Carrying out the integration over $x_g$ in Eq.\eqref{eq:amp}, one arrives at,
 \begin{align}
{\text{ Im}} {\cal M}\!=\!\frac{\vert \vec \epsilon_{\gamma \perp} \! \times \!  \vec l_{\rho \perp}\vert}{l_{\rho \perp}^2}\! \int \!\! dz {\cal T}\! (y,\xi,\xi,z) \phi_{\rho}(z) \frac{-i\pi H_g(\xi,\!\xi,\!t)}{2\xi}.
\end{align}
The hard-scattering amplitude $\mathcal{T}$ with $x_g=\xi$ is independent of $l_\rho$,
\begin{align}\label{eq:rho-a_T}
 {\cal T} (y,\xi,\xi,z) =&\frac{e C_\rho}{N_c}\frac{f_\rho}{f_a}\left[ -4\pi \alpha_s\frac{(1-2z)^2+4(1-z)zy}{2(1-z)^2z^2y}\right.\nonumber\\
 &\quad\left.+1.92\times 4\pi \alpha_e\frac{4y}{(1-z)z(1-y)} \right],
\end{align}
where $C_\rho=\frac{1}{\sqrt 2}( e_u-e_d)=\frac{1}{\sqrt{2}}$ is the flavor weight factor for the $\rho$ meson.  For $\omega$ and $\phi$, there are $C_\omega=\frac{1}{\sqrt 2}( e_u+e_d)=\frac{1}{3\sqrt{2}}$ and $C_\phi=e_s=-\frac{1}{3}$, and we can see that this factor suppressed the production $\omega$ or $\phi$ meson.

Note that the hard-scattering amplitude is divergent when $z$ approaches 0 or 1 which are the well known as end point singularities.
From a phenomenological standpoint, regularization is achievable by considering the transverse momentum dependence of the $\rho^0$ DA~\cite{Sun:2021gmi,Bhattacharya:2023hbq}. According to the general power counting rule, a simple scheme of introducing $p_\perp$ dependence is to  replace the lower and the upper integration limit of $z$ by ${\langle p_\perp^2 \rangle/l_{\rho \perp}^2}$ and ${1-\langle p_\perp^2 \rangle/l_{\rho \perp}^2}$, respectively. The factor $\langle p_\perp^2 \rangle $ is the mean squared transverse momentum  of the quark inside $\rho^0$ and in this work $\langle p_\perp^2 \rangle =(100\mathrm{MeV})^2 $ is chosen. More sophisticated treatment of end point singularities can be found in Refs.~\cite{Goloskokov:2005sd,Goloskokov:2006hr}. 

The initial photon is from the electromagnetic field of an incoming nucleus. For incoming heavy ions, the induced coherent photon distribution function can be readily computed in classical electrodynamics~\cite{Bertulani:1987tz,Bertulani:2005ru, Baltz:2007kq},
\begin{align}
x_\gamma f_\gamma(x_\gamma)= \frac{2 Z^2 \alpha_{\text{em}}}{\pi} &\bigg [  \zeta K_0(\zeta) K_1(\zeta) \nonumber\\ 
&-\frac{\zeta^2}{2} \left ( K_1^2(\zeta)-K_0^2(\zeta) \right ) \bigg],
\end{align}
where $x_\gamma$ is the longitudinal momentum fraction of a nucleon taken by the photon, $K_0(\zeta)$ and $K_1(\zeta)$ are the modified Bessel functions of the second kind, and $\zeta\equiv 2x_\gamma M_p R_A $ with $M_p$ and $R_A$ being  the proton mass and nuclear radius, respectively.  
If the incoming nucleus is a proton, we use the photon PDF computed in QED at the lowest order (see Ref.~\cite{Liu:2020rvc} and the reference therein),
\begin{align}
f_{\gamma}\left(x_{\gamma}, \mu\right)=\frac{\alpha_{e}}{2 \pi} \frac{1+\left(1-x_{\gamma}\right)^{2}}{x_{\gamma}} \ln \frac{\mu^{2}}{x_{\gamma}^{2} M_{p}^{2}},
\end{align}
where the scale $\mu^2$ is set to be $0.1~\mathrm{GeV}^{2}$ in this work.

Assembling all pieces together, the corresponding cross section in the ultraperipheral collision of nuclei $\mathcal{A}$ and $\mathcal{B}$ is
\begin{widetext}
\begin{align}
 \frac{d \sigma}{d^2 l_{\rho \perp}  dy_\rho  dy_a}=&\frac{A^2}{512 \pi^4 B}  \frac{1}{ l_{\rho \perp}^2} \frac{x_\gamma  f_\gamma(x_\gamma)}{(x_\gamma s)^2}  \left |\int \! dz {\cal T} (y,\xi,\xi,z) \phi_{\rho}(z) \pi\! \left [i -\frac{\pi}{2} \frac{\partial }{\partial \ln 2\xi}2\xi  \right ] \! G(2 \xi)   \right |^2
 + ( {\cal A} \leftrightarrow {\cal B}),
\label{cs}
\end{align}
\end{widetext}
where the transverse momentum of the axion has been integrated out for all available phase space.  The partonic center-of-mass energy $\sqrt{\hat s}$ is related to final-state particles transverse momenta by $\hat s=2x_\gamma \xi s={l_{\rho \perp}^2}/{y (1-y)}$,  where $\sqrt{s}$ is the center-of-mass energy per nucleon pair. The variables $x_\gamma$ and $\xi$ are constrained by external kinematic variables  $x_\gamma=l_{\rho \perp} \left (e^{y_\rho}+e^{y_a}\right )/\sqrt s  $ and $2 \xi=l_{\rho \perp} \left (e^{-y_\rho}+e^{-y_a}\right )/\sqrt s  $, where $y_\rho$, $y_a$ are the rapidities of the $\rho^0$ and the axion, respectively. Note that the double-slit-like interference effect is not included in this result~\cite{Klein:1999gv,Xing:2020hwh,Brandenburg:2022jgr}.
 
In experiments, the four-momentum of the recoiled ion cannot be directly measured at the LHC due to its extremely small scattering angles. 
So we look at events in which only the recoiled $\rho^0$ is detected with no other particle observed. 
The main decay channel of $\rho^0$ is $\pi^+ \pi^-$ which can be easily identified by the LHC detector.
To estimate the cross section, we numerically integrated the rapidities of $\rho^0$ and $a$ from -4 to 4.
The $\rho^0$ transverse momentum is integrated from $2$ to 20~GeV.
The total cross sections in lead-lead collisions and proton-proton collisions at LHC are,
 \begin{equation} 
 \sigma_{t} =\left\{
 \begin{aligned}
     1.99\times 10^{-1} \mu \text{b}\left(\frac{\text{10~TeV} }{f_a}\right)^2,\quad& \text{PbPb@5.2TeV},\\
     9.39\times 10^{-11}\mu b \left(\frac{\text{10~TeV} }{f_a}\right)^2, \quad &\text{pp@14 TeV}.
 \end{aligned}\right.
 \end{equation}
Because of the $Z^2A$ enhancement, the total cross section of mono-$\rho^0$ production in lead-lead hard exclusive collisions is much larger than that in $pp$ collisions.
Given the integrated luminosity $13~\text{nb}^{-1}$ of the heavy ion run of the LHC, 25 axion production events can be collected in UPCs for $f_a=100~\text{TeV}$. 

\section{ QCD background}\label{sec:background}

\begin{figure}[!htb]
\centering
\includegraphics[angle=0,scale=0.6]{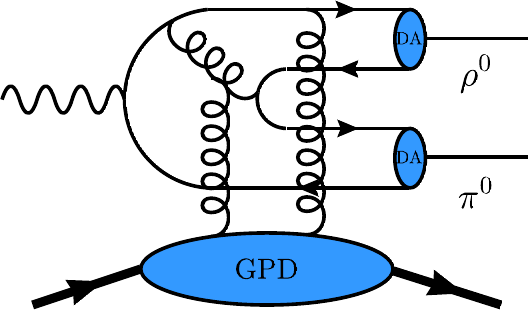}
\caption{A sample Feynman diagram contributing to exclusive $\rho^0+\pi^0$ production.} \label{rhopi}
\end{figure}

The apparent QCD background is $\gamma+ {\cal A} \rightarrow \rho^0+ {\cal A}$ by exchanging a $t$-channel Pomeron.
Nonetheless, this process does not contribute much to $\rho^0$ with transverse momentum larger than a few GeV due to the power law behavior($1/t^4$) in the large $t$ region~\cite{Ji:2003fw,Hoodbhoy:2003uu,Tong:2021ctu,Sun:2021gmi}.
For large transverse momentum $\rho^0$, the exclusive  $\rho^0$ and $\pi^0$ photoproduction  $\gamma+ {\cal A} \rightarrow \rho^0+ \pi^0+ {\cal A}'$ is the main source of the QCD background. The Feynman diagram for such process is given in Fig.~\ref{rhopi}. If both photons from  $\pi^0$ decay escape detection, this process could mimic the ALP signal. The exclusive meson pair production has been investigated to study the proton chiral-odd GPDs.~\cite{Ivanov:2002jj,ElBeiyad:2010pji,Siddikov:2023qbd,Siddikov:2022bku}. 
it has also been suggested that exclusive meson pair production in $e^+ e^-$ annihilation could provide useful constraints on the DAs of light vector mesons~\cite{Lu:2024cjp}. 

The cross section of meson pair production can be computed following a similar procedure. We first calculate the imaginary part of the amplitude and then obtain the approximate real part contribution using the dispersion relation as in Eq.\eqref{eq:dispersion}. One obtains
\begin{widetext}
   \begin{align}
\frac{d \sigma}{d^2 l_{\rho \perp}  dy_\rho  dy_{\pi}}=\frac{A^2}{512 \pi^4 B}  \frac{1}{l_{\rho \perp}^6} \frac{x_\gamma  f_\gamma(x_\gamma)}{(x_\gamma s)^2}   \left |\int \! dz {\cal T}_{\rho \pi} (y,\xi,\xi ,z_1,z_2) \phi_{\rho}(z_1) \phi_{\pi}(z_2) \pi\! \left [i -\frac{\pi}{2} \frac{\partial }{\partial \ln 2\xi}2\xi \right ] \! G(2 \xi)   \right |^2,
\label{csrhopi}
 \end{align}
 \end{widetext}
where $y_\pi$ is the rapidity of $\pi^0$ and $\phi_{\pi}(z_2)$ is the DA of $\pi^0$, defined as in Eq.~\eqref{eq:rho_DA} with $\rho^0$ replaced by $\pi^0$.
In our calculation, we use $\phi_\pi(z_2)=6z_2(1-z_2)$ and the pion decay constant is 131~MeV \cite{ParticleDataGroup:2022pth}. 
The hard-scattering amplitude is given by
\begin{eqnarray}
 {\cal T}_{\rho \pi} (y,\xi,\xi,z_1,z_2)=-
\frac{ef_{\pi}f_{\rho}\pi^{2}\alpha _{s}^{2}}{N_{c}^{2}}\frac{{\cal T}_{N}}{{\cal T}_{D}}(1-y)y, \label{coe}
\end{eqnarray}
where the functions ${\cal T}_{N}$ and ${\cal T}_{D}$ are complicated and are presented in the Appendix. The end point singularities are treated as discussed below Eq.~\eqref{eq:rho-a_T}.
Integrated over the rapidities from $-4$ to 4, and the $l_{\rho\perp}$ from 2 to 20 GeV, the total cross sections in lead-lead collisions and proton-proton collisions at LHC are,
 \begin{equation}
 \sigma_t=\left\{\begin{aligned}
     &1.2\times 10^2 \mu \text{b},&\quad & \text{PbPb @5.2 TeV},\\
     &4.5\times 10^{-6} \mu \text{b}, &\quad & \text{pp @14 TeV}.
 \end{aligned}\right.
 \end{equation}
Given the integrated luminosity $13~\text{nb}^{-1}$ of the heavy ion run of the LHC, $1.56 \times 10^6$ such double-meson production events could be collected in UPCs.
\begin{figure}[htpb]
    \centering
    \includegraphics[width=0.45\textwidth]{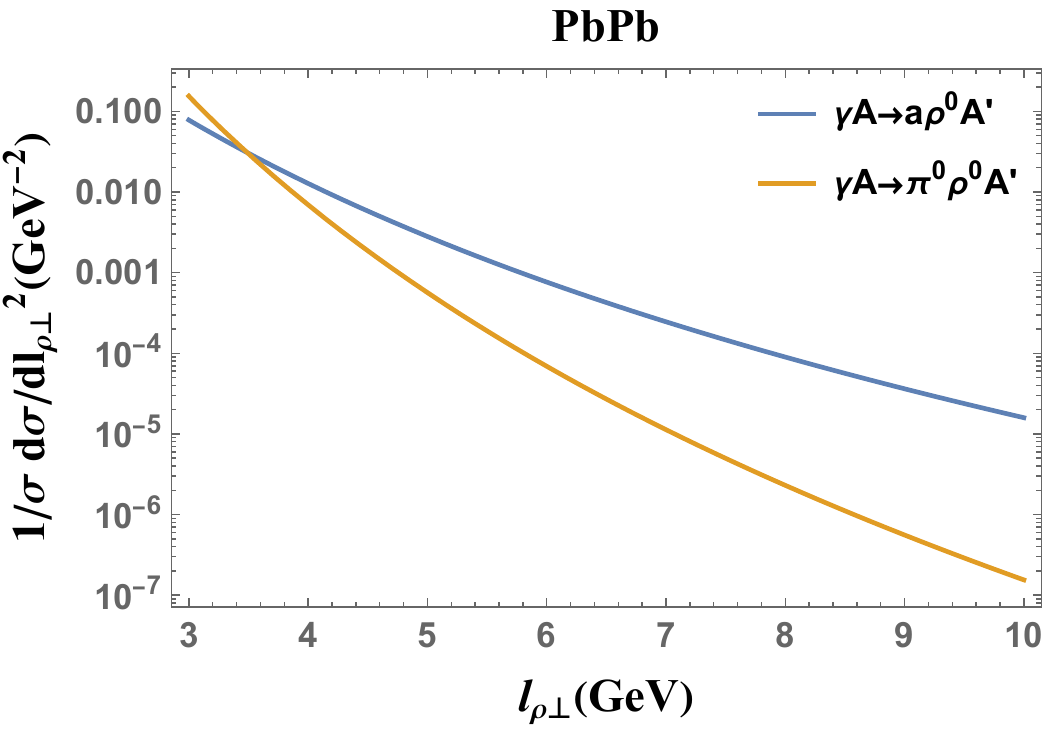} 
    \includegraphics[width=0.45\textwidth]{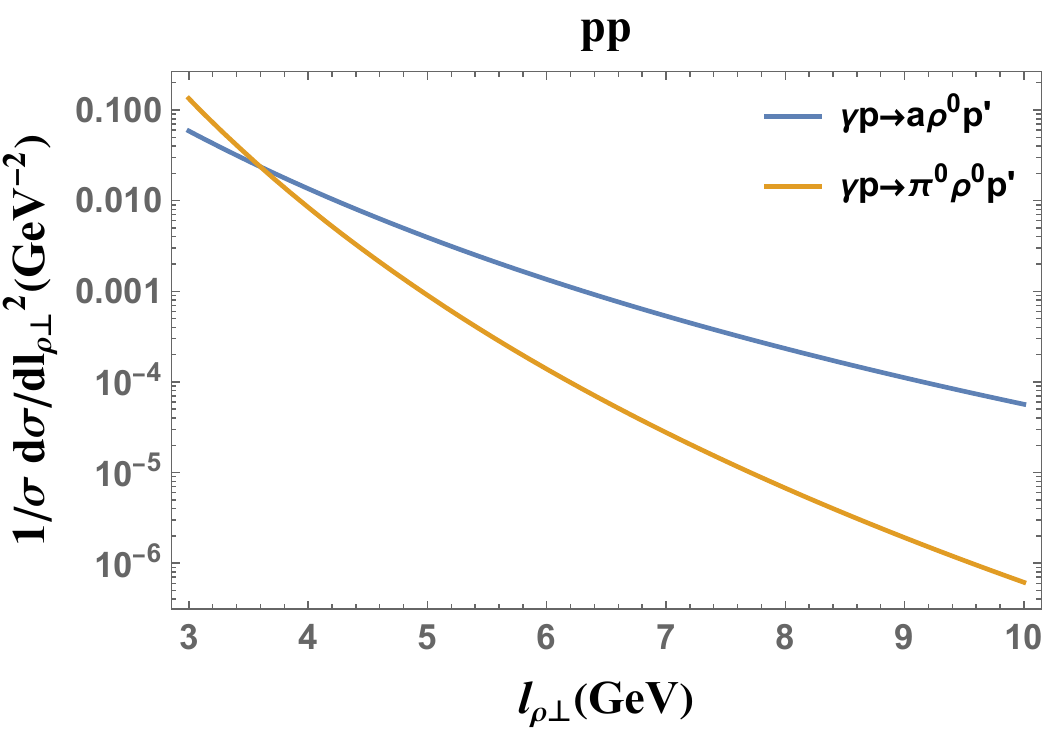}
    \caption{The normalized differential cross section of the signal process  $\rho^0+a$ photoproduction and the background process $ \rho^0 +\pi^0$ for $PbPb$ UPCs at $\sqrt s=5.2$~TeV (a) and for $pp$ hard exclusive collisions at $\sqrt s=14$~TeV (b) as functions of $l_{\rho \perp}$. The rapidity of $\rho^0$ is integrated over the range $[-4, 4]$. 
    }
    \label{average}
\end{figure}

This process contributes to the background only when both photons from $\pi^0$ decay are missed by the detector. 
In Fig.~\ref{average}, we compare the signal with the background processes, by normalizing the differential cross section of both respectively.
The rapidity is integrated from -4 to 4.    
The normalized differential cross sections are plotted as functions of the transverse momentum $l_{\rho \perp} $.
Notably, the QCD background decreases more rapidly than the axion production cross section. This behavior arises because the latter scales as ${l_{\rho \perp}^{-2}}$, while the former scales as ${l_{\rho\perp}^{-6} }$. Consequently, it is more efficient to collect signal events with relatively high transverse momentum.   

\section{Bounds on axion-gluon coupling at the LHC}\label{sec:constraint}

In this section, we investigate the prospect of searching for the long-lived axion through the mono-$\rho^0$ process at the LHC. 
Light axions decay dominantly to two photons \cite{Aloni:2018vki}.
The decay width could be calculated with the effective axion-photon coupling in Eq.\eqref{eq:effphoton},
\begin{align}
    \Gamma(a\to \gamma\gamma)=\frac{m_a^3 \vert c^{\mathrm{eff}}_{a\gamma\gamma}\vert^2}{4\pi f_a^2}. 
\end{align}
With the cancellation assumption given in Eq.\eqref{eq:cphoton}, $c^{\mathrm{eff}}_{a\gamma\gamma}$ is further suppressed by $m_a^2/m_\pi^2$ for $m_a\ll m_\pi$, making the decay width proportional to $m_a^{7}$.
Under this assumption, light axions are long-lived unless $m_a$ is close to the pion mass.

To estimate the sensitivity of the experiments, we introduce the likelihood function
\begin{align}
    L(n\vert b+s)=e^{-(b+s)} \frac{(b+s)^n}{n!},
\end{align}
where $n$, $s$, $b$ denote  the  numbers of observed events, the predicated mono-$\rho$ signal events and the SM background events, respectively. Consequently the significance is defined as
\begin{align}
    \mathcal{S}=\sqrt{-2\ln\frac{L(n\vert b+s)}{L(n\vert n)}}.
\end{align}
We further assume $n=b$ in order to obtain the expected limit on the signal. 
The theoretical uncertainties of both the signal and backgrounds are not included.
Then the exclusion at $95\%$ confidence level (CL). can be obtained from
\begin{align}
    \mathcal{S}=\sqrt{2\left(s+b\ln\frac{b}{b+s}\right)}\geq 2.
\end{align}
In the analysis, the background we considered is the $\rho^0+\pi^0$ production with both photons from the $\pi^0$ decay evading the detection. We refer to this possibility as $\epsilon_b$. Below we also define $\epsilon_s$ as the detection efficiency of the mono-$\rho^0$ signal events. 

\begin{figure}[htpb]
    \centering
    \includegraphics[width=0.45\textwidth]{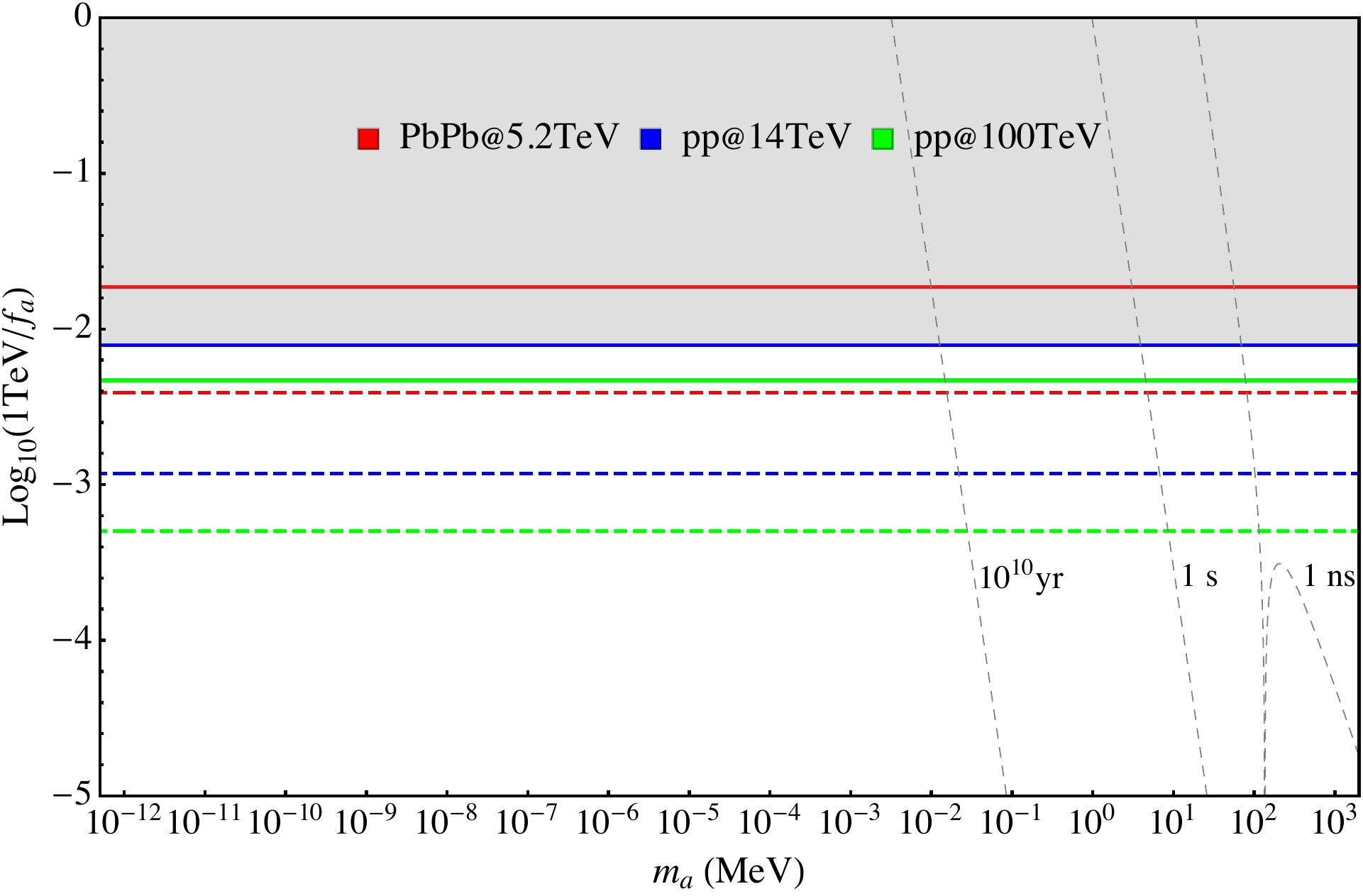}
    \caption{Constraints on the axion-gluon coupling  derived from the missing $k_t$ approach at the LHC. The red solid line represents the result from $Pb Pb$ collisions at $\sqrt{s}=5.2$~TeV with   $13~\text{nb}^{-1}$ of the integrated luminosity, while the blue (green) solid line corresponds to  the exclusion limit obtained  from $\sqrt{s}=14$~TeV (100~TeV)  $pp$ collisions with    $3000~\text{fb}^{-1}$ of the integrated luminosity.}
    \label{fig:exclusion}
\end{figure}

For axions with mass $m_a$ lighter than a few hundred MeV, their impact on the mono-$\rho^0$ kinematics at the LHC is negligible, resulting in the fact that the exclusion limits are not dependent on $m_a$.
Figure.~\ref{fig:exclusion}  illustrates the limits in  both the ideal and realistic scenarios.  
The bounds in the ideal case with $\epsilon_s=1$ and $\epsilon_b=0$ are displayed with dashed lines.
The limit on $f_a$ from the mono-$\rho^0$ process is as large as 260 TeV in 5.2 TeV lead-lead collisions with the luminosity of 13~nb$^{-1}$ and 850 TeV in 14 TeV proton-proton collision with the luminosity of 3000~fb$^{-1}$. It can even reach an impressive value of 2000~TeV on the future 100~TeV $pp$ colliders\cite{CEPCStudyGroup:2018ghi,FCC:2018vvp} with the luminosity of 3000~fb$^{-1}$.

The solid lines in Fig.~\ref{fig:exclusion} represent the results for a more realistic scenario with $\epsilon_s=0.99$ and $\epsilon_b=0.1\%$. We introduce the cut on the transverse moment of the $\rho$ meson as $3.5 \leq l_{\rho\perp}\leq 20~\mathrm{GeV}$, 
since the signal-to-background ratio is larger in the region of higher $l_{\rho\perp}$ as shown in Fig.~\ref{average}. The expected constraints on $f_a$ in this scenario are about 50~TeV in 5.2~TeV lead-lead collisions with the luminosity of 13~nb$^{-1}$, 130~TeV in 14~TeV proton-proton collision with the luminosity of 3000~fb$^{-1}$, and 210~TeV on the future 100~TeV proton-proton colliders with the luminosity of 3000~fb$^{-1}$. 

We also verify the assumption of long-lived axions as in Eq.\eqref{eq:cphoton}. In Fig.~\ref{fig:exclusion}, the gray dashed curves represent the parameters for axions with lifetimes $10^{10}$~yr, 1~s and 1~ns, respectively. Since the axion produced at LHC is highly boosted, the lifetime greater than 1~ns is sufficient for the axion to escape from the detector, requiring the axion mass to be less than 100~MeV. Nonetheless, if axions are dark matter, their lifetime should be comparable to the Universe age~$\sim 10^{10} $yr, necessitating the axion mass to be less than 0.1 MeV.

The constraint on the axion-gluon coupling can also be determined through monojet searching at the LHC.
With the integrated luminosity of 139~fb$^{-1}$, the ATLAS Collaboration obtains a limit on the $f_a$~\cite{ATLAS:2021kxv},
\begin{align}
    1/f_a< 8\times 10^{-3}~\mathrm{TeV}^{-1},\quad\text{at 95\% CL},
\end{align}
where we take $C_{agg}=1$ as in Eq.~\eqref{eq:L}.
The excluded region is presented as the gray region in Fig.~\ref{fig:exclusion}.

The axion-gluon coupling can also be studied indirectly through nonperturbative effects in chiral perturbation theory. A large parameter space for the axion can be constrained by intensity frontier experiments, such as beam dump experiments~\cite{CHARM:1985anb,Dobrich:2015jyk,FASER:2018eoc}, rare decays of kaon and pion~\cite{NA64:2020qwq,NA62:2020pwi,NA62:2021zjw,NA62:2023olg,E949:2005qiy,Gori:2020xvq,Dolan:2017osp}, and others; and the light axion can also be constrained from astrophysical and cosmological observations~\cite{Cadamuro:2011fd,Millea:2015qra,Ertas:2020xcc,Chang:2018rso,Depta:2020wmr}. Nevertheless, the contribution from axion-gluon coupling in such processes is involved with the axion-quark and axion-photon interactions. In this work, we investigate the axion-gluon coupling at high-energy scales, and the $agg$ vertex is free of nonperturbation. Furthermore, the contributions from axion-quark couplings are significantly suppressed due to the light quark masses. Consequently, the monohadron process in addition to the monojet process can provide direct constraints on the axion-gluon coupling at LHC and future high-energy colliders.

\begin{figure}[htpb]
    \centering
\includegraphics[width=0.45\textwidth]{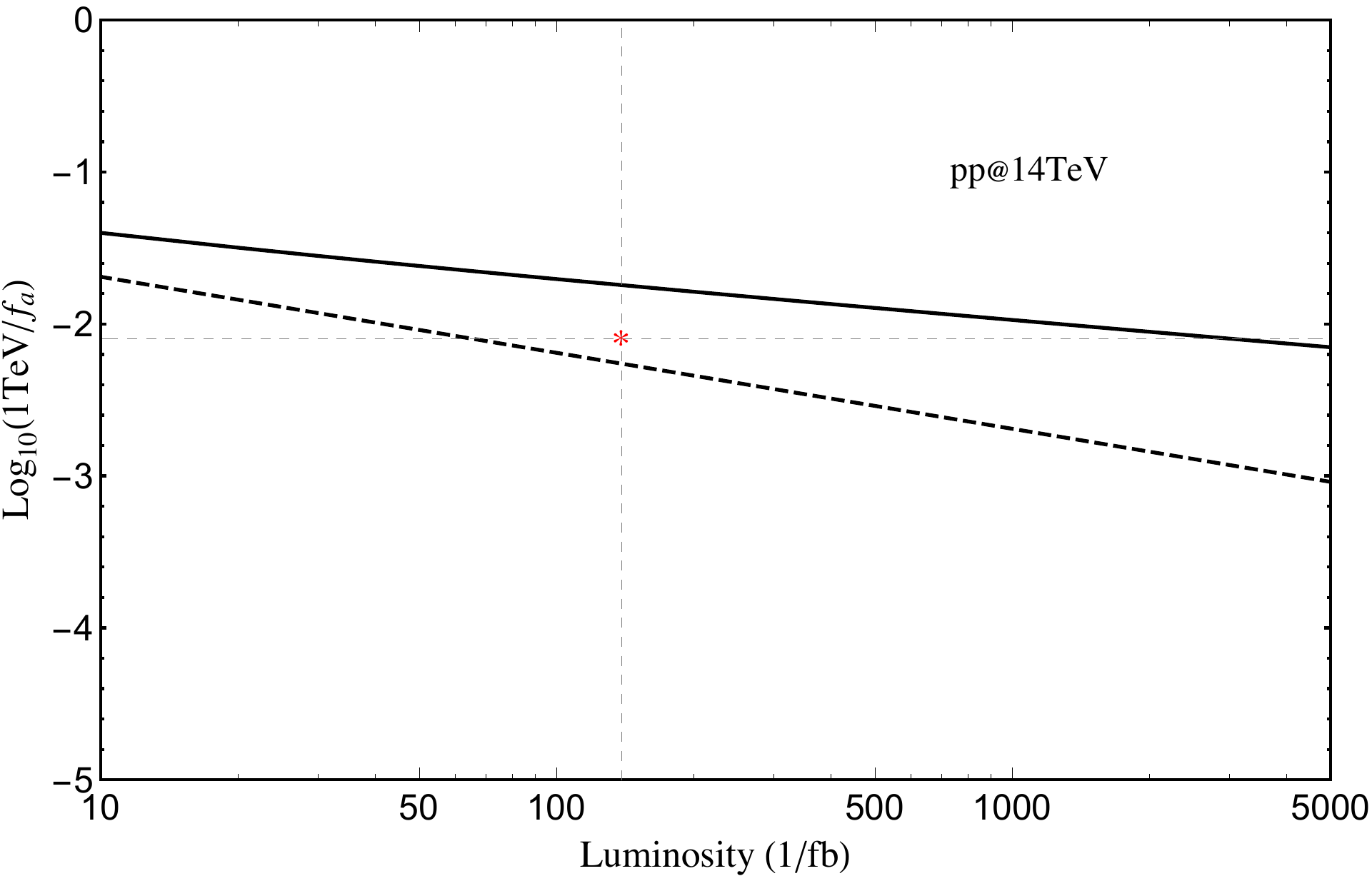}
    \caption{Constraints on the axion-gluon coupling through mono-$\rho$ and monojet in $pp$ collisions at the LHC. The solid line represents the mono-$\rho$ result with $\epsilon_s=0.99$ and $\epsilon_b=0.1\%$, while the dashed line corresponds to the result assuming perfect detection efficiency. The red star represents the constraint obtained with the monojet approach  by the ATLAS Collaboration~\cite{ATLAS:2021kxv}.}
    \label{fig:14tev}
\end{figure}

Figure.~\ref{fig:14tev} further shows the expected limit of the $f_a$ through mono-$\rho^0$ searching in $14$ TeV $p p$ collision as a function of the luminosity. The solid line is the result obtained with $\epsilon_s=0.99$ and $\epsilon_b=0.1\%$, while the dashed line represents the result with perfect detection efficiencies. The star denotes the limit obtained by ATLAS through monojet searching with the luminosity of 139~fb$^{-1}$. We observe that the sensitivity of the mono-$\rho$ process is comparable to that of the monojet process. It is important to note that the analysis could be expanded to include the production of other hadrons, which could improve the sensitivity.

\section{Summary}\label{sec:summary}

In this work, we investigate the feasibility of constraining the axion properties via exclusive  processes, in which the initial particles remain intact after interaction. We take the associated production of an axion with a $\rho^0$ meson in photon-nucleus and photon-proton collisions as examples. With the axion escaped from the detector, the final state presents a monohadron signature. The key advantage of exclusive processes lies in the simple and clean background. Specifically, the SM background for this process involves a high boosted $\rho^0$ meson, which are mainly from the production of $\rho^0$ and $\pi^0$, with neither photons from the $\pi^0$ decay observed due to limited detector efficiency.

Compared to the inclusive production of axions, the diffractive process introduces a complication related to the specific hadrons in the final state. In our example, there are  $\rho^0$ and $\pi^0$ mesons that appear in the signal and background processes. Their production rates have been well described with the distribution amplitudes used in this work. More sophisticated  treatments are also feasible. Given the more accumulated data and our improved understanding of QCD factorization, the uncertainties arising from the final-state hadrons shall become negligible.

In this work, we assume that the $\rho^0+\pi^0$ production events   with missing $\pi^0$ constitute  0.1\% fraction  of the total production rate. After applying a cut on the $\rho^0$ transverse momentum $3.5\leq l_{\rho\bot}\leq 20\mathrm{GeV}$, we have determined the bounds on the axion decay constant $f_a$ to be 50~TeV for 5.2~TeV $PbPb$ UPCs, and 130~TeV for 14~TeV $pp$ hard exclusive collisions. With future 100~TeV $pp$ colliders, the bound on $f_a$ can reach 210~TeV. Although the larger photon and gluon fluxes in the UPC process, the $pp$ hard exclusive process provides a more stringent bound due to a much larger integrated luminosity. These bounds we obtained are comparable to those derived from monojet events, making the monohadron approach complementary to the existing method in directly constraining the axion-gluon coupling.
 
There are several directions in which the current analysis can be refined and extended. In this work, we only consider the simplest cut on the $\rho^0$ transverse momentum, while the more sophisticated cut strategy on the final-state kinematics can be used to further raise the signal-to-background ratio.
In addition, one can improve the accuracy of perturbative calculation by including the saturation effect and using the more sophisticated models of GPDs. 
Moreover, one may consider different species of light meson production accompanied by an emission of low-mass axions.
Finally, detectors with better performance could largely improve the exclusion bound by suppressing the SM background.

\ 
 
\begin{acknowledgments}
We thank Bing-xuan Liu, Tian-bo Liu, Yu-ming Wang, Chi Yang, Shuai Yang and Jin-long Zhang for helpful discussions.  J. Z. has been supported by the National Science Foundations of China under Grant No.\ 12175118 and 1232100005 and the National Science Foundation under Contract No.~PHY-1516088. 
S. S. B. and H. Z. are supported by the National Natural Science Foundation of China (NSFC) under Grant No.~12075136 and the Natural Science Foundation of Shandong Province under Grant No.~ZR2020MA094.
\end{acknowledgments}

\appendix

\section{The hard coefficients in Eq.~\eqref{coe}}
 
The hard coefficients in Eq.~\eqref{coe} are given by,
\begin{widetext}
\begin{eqnarray}
   {\cal T}_{N}\!\!&=&-\left(z_1-1\right) z_1 \left(z_2-1\right){}^2 z_2^2 \left(8 \left(2 z_1-1\right) z_2^3+8 \left(z_1^2-4 z_1+2\right) z_2^2+\left(-10 z_1^3+7 z_1^2+11 z_1-8\right) z_2+5 \left(z_1-1\right) z_1^2\right)\notag\\&&+(1-y) \left(z_2-1\right) z_2 \left(4 \left(8 z_1^3-12 z_1^2+2 z_1+1\right) z_2^5+2 \left(2 z_1^4-44 z_1^3+51 z_1^2+z_1-8\right) z_2^4\notag\right.\\&&\left.-2 \left(10 z_1^5-21 z_1^4-8 z_1^3+7 z_1^2+24 z_1-12\right) z_2^3+\left(12 z_1^6-6 z_1^5-33 z_1^4+56 z_1^3-75 z_1^2+62 z_1-16\right) z_2^2\notag\right.\\&&\left.-\left(z_1-1\right){}^2 \left(12 z_1^4-6 z_1^3-z_1^2+20 z_1-4\right) z_2+\left(z_1-1\right){}^3 z_1 \left(5 z_1^2-2 z_1-4\right)\right)\notag\\&&+(1-y)^2 \left(-8 \left(2 z_1^3-3 z_1^2-z_1+1\right) z_2^7+\left(4 z_1^4+48 z_1^3-29 z_1^2-79 z_1+35\right) z_2^6\notag\right.\\&&\left.-\left(12 z_1^4+4 z_1^3+123 z_1^2-217 z_1+60\right) z_2^5+\left(-4 z_1^6+12 z_1^5+9 z_1^4-108 z_1^3+306 z_1^2-270 z_1+50\right) z_2^4\notag\right.\\&&\left.+2 \left(8 z_1^7-24 z_1^6+32 z_1^5-39 z_1^4+91 z_1^3-147 z_1^2+89 z_1-10\right) z_2^3\notag\right.\\&&\left.-\left(z_1-1\right){}^2 \left(24 z_1^5-56 z_1^4+56 z_1^3-46 z_1^2+61 z_1-3\right) z_2^2\notag\right.\\&&\left.+\left(z_1-1\right){}^3 z_1 \left(16 z_1^3-32 z_1^2+24 z_1-13\right) z_2-4 \left(z_1-2\right) \left(z_1-1\right){}^4 z_1^2\right)\notag\\&&-(1-y)^3 \left(z_1+z_2-1\right){}^2 \left(\left(8 z_1-4\right) z_2^5+\left(9 z_1^2-29 z_1+1\right) z_2^4+\left(-6 z_1^3-9 z_1^2+31 z_1+10\right) z_2^3\notag\right.\\&&\left.+\left(9 z_1^4-9 z_1^3+18 z_1^2-22 z_1-7\right) z_2^2+z_1 \left(8 z_1^4-29 z_1^3+31 z_1^2-22 z_1+12\right) z_2-\left(z_1-1\right){}^2 z_1^2 \left(4 z_1-11\right)\right)\notag\\&&-(1-y)^4 \left(z_1+z_2-1\right){}^3 \left(7 z_2^3-7 \left(z_1+1\right) z_2^2+z_1 \left(2 z_1+5\right) z_2-2 \left(z_1-1\right) z_1^2\right)\notag\\&&+(1-y)^5 \left(z_2-z_1\right){}^2 \left(z_1+z_2-1\right){}^4,
\end{eqnarray}
and,
\begin{eqnarray}
  {\cal T}_{D}&=&9 \left(z_1-1\right){}^2 z_1^2 \left(z_2-1\right){}^2 z_2^2 \left(z_1 z_2-(1-y) \left(z_1+z_2-1\right)\right){}^2 \left((1-y) \left(z_1+z_2-1\right)+\left(z_1-1\right) \left(z_2-1\right)\right){}^2.
\end{eqnarray}
\end{widetext}


\end{document}